# Beyond intensity modulation: new approaches to pump-probe microscopy

JUN JIANG,[1,†] DAVID GRASS,[2,†] YUE ZHOU,[2] WARREN S. WARREN,[1,2,3,4] MARTIN C. FISCHER[2,3,*]

[1]*Department of Biomedical Engineering, Duke University, Durham, North Carolina, 27708, USA*
[2]*Department of Chemistry, Duke University, Durham, North Carolina, 27708, USA*
[3]*Department of Physics, Duke University, Durham, North Carolina, 27708, USA*
[4]*Department of Radiology, Duke University, Durham, North Carolina, 27710, USA*
†*J.J. and D.G. contributed equally to this work*
*\*Corresponding author: martin.fischer@duke.edu*

**Pump-probe microscopy is an emerging nonlinear imaging technique based on high repetition rate lasers and fast intensity modulation. Here we present new methods for pump-probe microscopy that keep the beam intensity constant and instead modulate the inter-pulse time delay, the relative polarization, or the pulse length. These techniques can improve image quality for samples that have poor heat dissipation or long-lived radiative states, and can selectively address nonlinear interactions in the sample. We experimentally demonstrate this approach and point out the advantages over conventional intensity modulation.**

Pump-probe microscopy is the combination of pump-probe spectroscopy and laser scanning microscopy [1]. It falls into the broad class of nonlinear microscopy and accesses contrast mechanisms such as two-photon absorption, excited state absorption, stimulated emission, ground state bleaching and stimulated Raman scattering. Pump-probe microscopes not only provide image contrast, but also resolve the ultrafast excited state dynamics of the sample. The combination of spatial and temporal dimension makes pump-probe microscopy a powerful technique with high molecular and chemical specificity. The broad range of nonlinear optical processes provides molecular contrast without the need for exogenous dyes and labels, and thus preserves the local biochemical environment of the target molecules.

In a conventional pump-probe experiment, a pump pulse creates excited state populations in the sample. After a variable delay, the populations are probed by a probe pulse. Pump-probe measurements depend strongly on the transfer of population between the involved states and through this dependence provide molecular specificity. Current pump-probe microscopes and spectroscopy setups commonly utilize an intensity modulation transfer scheme [1]. In this method, the pump intensity is repeatedly switched on and off, and the probe absorption is measured with and without a pump pulse present, as a function of a variable inter-pulse delay $\Delta t$. The pump-induced change in absorption of the sample is reflected in the power difference between two consecutive probe measurements. This is done for many repetitions by intensity modulating the pump pulse train with a frequency $\Omega_{mod}$ and performing a lock-in measurement of the probe pulse train at the same frequency. This modulation transfer detection measures the average difference between both probe powers $\Delta A(\Delta t) = \langle P_{pr}(P_{pu}) - P_{pr}(P_{pu} = 0) \rangle (\Delta t)$ as a function of time delay between pump and probe pulse over many probe pulse repetitions. Especially for imaging applications, the intensity of the pump beam is modulated at high frequency (typically >1 MHz) to avoid technical low frequency noise (such as 1/f electronic noise, classical laser intensity noise) and to achieve reasonable imaging speed when pump and probe beam are raster-scanned across the sample. This method commonly measures pump-induced probe changes on the parts per million level [1].

Although intensity modulation is a very sensitive method to detect weak nonlinear signals, it has limitations. Typically, the pump pulses excite several molecular states at the same time, which results in complex (hence molecularly specific) pump-probe delay curves. However, spurious signals can also be generated: absorption in the sample may generate heat and modulation of the pump intensity generates local and time-dependent thermal expansion and refractive index changes. While this mechanism is taken advantage of in photothermal imaging [2], it is generally undesired in pump-probe imaging. Spurious signals can also result from the emission of radiative states with lifetimes much longer than the repetition rate of the laser or from pump-induced fluorescence at the probe wavelength. For example, autofluorescence exists in many samples and can leak into the detection channel if the spectrum of fluorescence emission overlaps with the spectrum of the probe beam. This pump-induced fluorescence background signal is independent of inter-pulse time delay and can be separately imaged and subtracted from the pump-probe image in post-processing, but subtraction will fail when the sample is not stationary (e.g., due to blood flow, in microfluidics, or from vibrations due to breathing or heart beating, etc.). Problems like thermal lensing and background signals also arise in stimulated Raman scattering microscopy for which various methods such as frequency modulation and stimulated Raman gain and opposite loss detection have been demonstrated to suppress background signals [3-6]. However, some of these methods introduce a third wavelength that is different from the original pump and probe (or Stokes) beam, which could cause additional transient processes like fluorescence emissions, and would add additional layers of complexity to the microscope. In a similar approach, intensity modulation of two phase shifted pump pulse trains for pump-probe measurements of blood oxygenation have been shown [7].

In this work, we demonstrate an enhanced modulation approach for pump-probe microscopy, in which the pump beam intensity can be held constant, while simultaneously modulating the inter-pulse time delay, the relative polarization, or the pulse length. This method allows suppression of spurious signals from emissive processes at the probe wavelength and thermal effects. In addition, it allows for selective imaging of molecular processes, e. g. based on their polarization, providing a powerful tool to increase chemical specificity of pump-probe microscopes.

On a fundamental level, the operation performed by the detection system in a pump-probe microscope is simple: a subtraction of probe powers conditioned on the presence of a pump pulse. Our scheme generalizes this operation: instead of simply switching the pump pulse on and off, the pump pulse is switched between two different pump states with adjustable power, polarization and temporal duration. The measured signal has the general form of $\Delta A(\Delta t) = \langle P_{pr}(a_0 P_{pu}(\beta_0, \tau_0, \Delta t_0)) - P_{pr}(a_1 P_{pu}(\beta_1, \tau_1, \Delta t_1)) \rangle$, with $a_0 (a_1)$ denoting the pulse amplitudes of the first (second) pump pulse, $\beta_0 (\beta_1)$ the polarization angle with respect to the probe polarization, $\tau_0 (\tau_1)$ the pulse length and $\Delta t_0 (\Delta t_1)$ the time delay with respect to the probe pulse, as illustrated in Fig. 1(b). This generalization can be used to implement multiple pump-probe modalities. Conventional intensity modulation is recovered in the special case of $a_0 = 0$ and varying $\Delta t_1$ to acquire a transient absorption curve.

Time-delay modulation can be used to reduce thermal effects and fluorescence backgrounds of radiative states. Instead of switching the pump on and off, the pump power is kept constant ($a_0 = a_1$) and the inter-pulse delay is modulated [7-9]. Another imaging modality, polarization modulation, can be used to selectively enhance or suppress nonlinear processes that are sensitive to the relative polarization angle between pump and probe. Modulating the state of polarization has already been used in pump-probe spectroscopy, but previous approaches are not suitable for imaging, for example due to bandwidth limitations [10, 11]. Finally, pulse-width modulation can be achieved by alternating between pump pulses of different duration.

The experimental setup is sketched in Fig. 1(a) and similar to [12]. We derive a pump and a probe pulse train from an ultra-broadband laser source (Coherent, Vitara UBB) via spectral filtering. The pump pulse train passes through an acousto-optical modulator (AOM) which separates the pump into the 0th and 1st diffraction order. The 1st order passes a delay line (pump delay line) and is superimposed with the 0th order on a beam splitter, resulting in the pump laser beam. An optional dispersive element (glass) in the 1st diffraction order beam allows for dispersion control of the pulse width $\tau_1$ and a waveplate ($\lambda/2$ plate) for control of the polarization angle $\beta_1$ of the 1st diffraction order of the pump beam. A second delay line (probe delay line) is used to introduce a time delay between pump and probe pulse train. A combination of both delay lines allows for independent control of the time delay between 0th order pump and probe $\Delta t_0$ and time delay between 1st order pump and probe $\Delta t_1$, see Fig. 1(b). Changing the relative power between 0th and 1st order beam is used to control the parameters $a_0$ and $a_1$. After spatially overlapping pump and probe, they are sent into a conventional laser scanning microscope. Before detection of the probe pulse train with a photodiode, the pump pulse train is rejected with a dielectric filter.

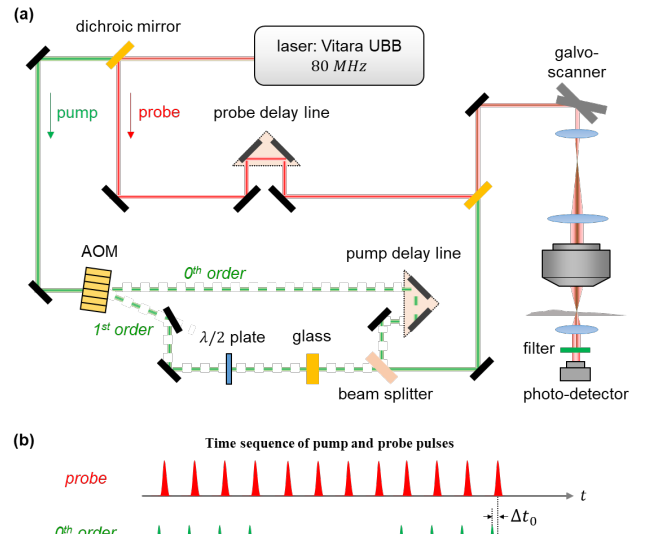

Fig. 1 (a) Experimental setup: a pulsed laser (Coherent, Vitara UBB) is split into a pump (green line) and a probe (red line) pulse train. The pump is further split by an acousto-optical modulator (AOM) into the 0th and 1st diffraction order. After passing an optional dispersive element (glass) and a half-wave plate ($\lambda/2$ plate), the diffraction orders are recombined and superimposed with the probe beam. The resulting beam is sent into a laser scanning microscope. (b) Pulse schematic: The two delay lines are used to control the delay $\Delta t_0$ between probe and 0th order and the delay $\Delta t_1$ between probe and 1st order. The AOM operates at frequency $\Omega_{mod}$, switching the pump pulse train between 0th and 1st diffraction order.

In a first experiment, we demonstrated time-delay modulation to suppress thermal lensing and spurious radiative signals. The laser system operates at a repetition rate of 80 MHz, separating pulses by 12.5 ns. Molecular processes with lifetimes significantly longer than 12.5 ns will appear as a constant offset in transient absorption curves, as these states cannot relax to equilibrium between pulses; similarly will spontaneous fluorescence emission at the probe wavelength. For time-delay modulation, the microscope is set to record $\Delta A(\Delta t_1 = \Delta t)$, with $a_0 = a_1$ (meaning both pump pulses have the same power) and $\Delta t_0$ such that the first pump pulse arrives after the probe pulse (in our experiments by 8 ps). Therefore, the first pump pulse does not influence the ultrafast dynamics measured by the probe (in ultrafast pump-probe imaging, the dynamics of interest are typically on the order of <100 ps, much faster than the temporal separation of pulses from the laser). The arrival time of the second pump pulse is scanned to measure transient absorption. Spurious signals with lifetimes much longer than the pulse repetition time and spontaneous fluorescence are present in both cases, whether the pump arrives before or after the probe pulse. The detection system measures the difference between both probe pulses and thereby eliminates these signals. We experimentally demonstrate that time-delay modulation recovers similar pump-probe images and transient absorption dynamics as ones acquired with conventional intensity modulation while, eliminating detrimental signals.

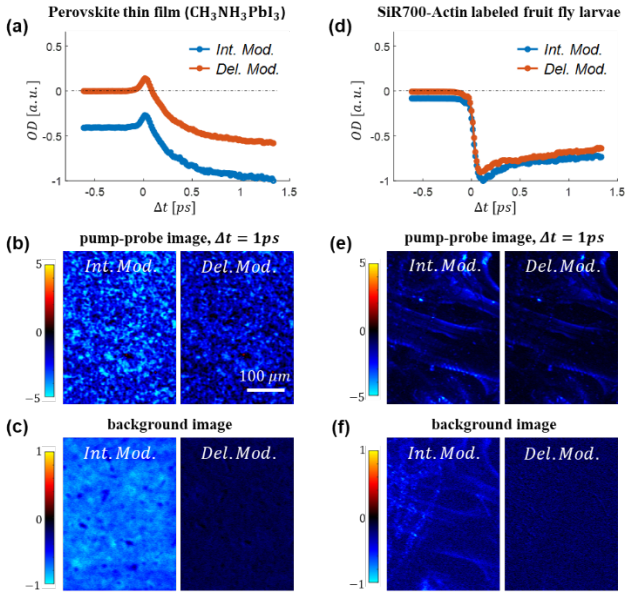

Fig. 2 Transient absorption curve comparison between intensity modulation (blue curves) and time-delay modulation (red curves) for a perovskite thin film (a) and a fruit fly larva (d), spatially averaged over the region of interest shown in (b) and (e), respectively. The negative offset in the intensity modulation images (red curves) originates from long-lived fluorescent states which are intrinsically eliminated in the time-delay modulation scheme. Pump-probe image comparison between intensity-modulation and time modulation for perovskite thin film (b) and fruit fly larva (e). While the signal images are similar, the background images (c) and (f), acquired at negative time delay, contain structure which pollutes the intensity modulation signal image.

We chose two samples: a perovskite thin film ($CH_3NH_3PbI_3$) and a dehydrated fruit fly larva labeled with SiR700-Actin. Perovskites are promising next generation solar cell materials and are currently heavily investigated with pump-probe microscopy to help unravel the ultrafast charge carrier dynamics [13, 14]. The fruit fly larva is a common biological sample. Fig. 2(a) and Fig. 2(d) show the transient absorption dynamics of the perovskite thin film and the fruit fly larva sample, acquired with intensity modulation (blue curves) and with time-delay modulation (red curves). Both samples were imaged with a center pump wavelength of 720 nm and a center probe wavelength of 817 nm. The curves are spatial averages of the images shown in Fig. 2(b) and Fig. 2(e), respectively. The blue and red curves, besides a constant offset, show the same ultrafast transient absorption dynamics. The large negative offset in the intensity modulation images originates from long-lived fluorescent emission that passes through the pump rejection filter. This offset is eliminated in the time-delay modulation scheme. The advantage becomes even more eminent by inspecting the images rather than the average curves. Fig. 2(b) and Fig. 2(e) show pump-probe images for intensity and time delay modulation of both samples at a time delay of $\Delta t = 1 ps$. The actual images are almost identical, but the background images shown in Fig. 2(c) and Fig. 2(f), acquired at a negative time delay of $\Delta t = -8 ps$, contain a lot of structural information that is hidden beneath the transient absorption image. In the case of time-delay modulation, this spurious background signal is eliminated. While post-processing can be used to subtract the background from the transient absorption image for a stationary sample, subtraction fails for non-stationary samples.

In the next experiment, we demonstrate that we can selectively image chemical species based on their polarization-dependent pump-probe signals. As many contrast mechanisms in pump-probe microscopy have a distinct polarization dependence, polarization modulation pump-probe microscopy increases chemical contrast for a variety of specimens. Here we consider a simple dipole transition moment model (see supplementary information) to describe the polarization dependence of nonlinear interactions. For example, in a ground state bleaching (GSB) event, pump and probe pulses compete for the same molecular transition and the contrast is stronger if pump and probe have the same polarization. For an excited state absorption (ESA) event, the pump excites population from the ground state to an excited state, and the probe pulse further excites the population to a higher energy state. In case of fast excited state depolarization or heterogeneous ensembles we can neglect the polarization dependence of excited state absorption. Generally, the polarization dependence is given by $s_x = a_x + b_x Cos[2(\beta - \beta_x)]$ with $\beta = \beta_0 - \beta_1$ the polarization angle between both pump pulses, $a_x$, $b_x$ and $\beta_x$ parameter depending on the nonlinear process and dipole orientations. For the simple GSB (ESA) process described above, one would find $a_{gsb} = 2/3$ ($a_{esa} = 1$), $b_{gsb} = 1/3$ ($b_{esa} = 0$) and $\beta_x = 0$. In a general approach, the polarization angle $\beta$ can be heuristically adjusted to increase the contrast in pump probe images and to differentiate chemical species. In this sense, polarization modulation is an additional contrast mechanism for pump-probe microscopy, just as phase contrast imaging is an additional contrast mechanism for conventional light microscopy.

If the polarization dependence of a chemical species is known *a priori*, polarization-based pump-probe can be used to selectively enhance or suppress the signal of this species during imaging. For example, if a sample contains a mix of GSB and ESA active species, a measured pump-probe image contains a superposition of both signals $m_{\Delta t}(\beta) = s_{gsb}(\beta)g_0(\Delta t) + s_{esa}(\beta)e_0(\Delta t)$, with $g_0(\Delta t)$ and $e_0(\Delta t)$ the pump-probe signal as a function of time delay for the GSB and ESA process, respectively. If we measure the sample with parallel $m_{\Delta t}(\beta = 0)$ and perpendicular $m_{\Delta t}(\beta = 90°)$ polarization between pump and probe, it allows us to reconstruct the individual molecular signals via:

$$g_0(\Delta t) = c_1 m_{\Delta t}(0) - c_2 m_{\Delta t}(90) \qquad (1)$$

$$e_0(\Delta t) = c_3 m_{\Delta t}(0) - c_4 m_{\Delta t}(90) \qquad (2)$$

with $c_i$ coefficients that solely depend on the polarization dependence of the individual components (see supplementary information). In a conventional intensity modulation scheme, this would mean at least two pump-probe images need to be recorded, one for each polarization setting to extract either GSB or ESA signal (or both). With polarization modulation, the GSB signal can be directly imaged as $g_0 = \left\langle P_{pr}\left(c_1 P_{pu}(\beta = 0) - \right.\right.$

$P_{pr}\left(c_2 P_{pu}(\beta = 90)\right)\right\rangle$ and the ESA signal as $e_0 = \left\langle P_{pr}\left(c_3 P_{pu}(\beta = 0)\right) - P_{pr}\left(c_4 P_{pu}(\beta = 90)\right)\right\rangle$. The intensity and polarization degrees of freedom allow us to directly measure either the GSB signal $g_0(\Delta t)$ or the ESA signal $e_0(\Delta t)$ by a acquiring only a single pump-probe image.

For demonstration purposes, we created a sample with two distinct chemical species that have clean and simple pump-probe dynamics. We choose grains of Lapis Lazuli, a historical blue pigment used for painting as a representative for a GSB process, and a powder form of Nile Red, an intracellular lipid stain as a representative for an ESA process. Besides availability, we picked these chemicals because of their distinct polarization dependence. Both were independently characterized (see supplementary information) and used to calculate the parameters $c_1, c_2, c_3$, and $c_4$.

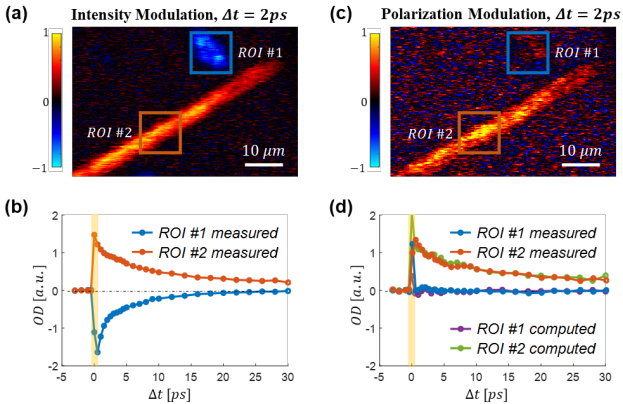

Fig. 3: (a) Intensity modulation and (c) polarization modulation image at $\Delta t = 2$ ps of Lapis Lazuli (blue box, ROI1) and Nile Red (red box, ROI2). (b) Intensity-modulation pump-probe curves averaged over ROI1 (blue) and ROI2 (red). (d) Polarization-modulation pump-probe curves; green and purple curves represent computed results based on intensity modulation, the red and blue curves are measured signals. Polarization modulation clearly suppresses the GSB signal in ROI1 and agrees with computed result.

We imaged the sample consisting of Lapis Lazuli and Nile Red with a center pump wavelength of 720 nm and a center probe wavelength of 817 nm. A conventional intensity modulation measurement with parallel polarization between pump and probe is shown in Fig. 3(a) at a time delay of $\Delta t = 2$ ps. Fig. 3(b) shows pump-probe curves of ROI1 (blue, Lapis Lazuli) and ROI2 (red, Nile Red). Both signals are clearly visible in the image as well as in the time delay traces. Fig. 3(c) shows the same field of view imaged with polarization modulation, set up to only image the ESA process (Nile Red). The 1st order pump beam is chosen to have a polarization angle of $\beta = 90°$ with respect to the 0th order pump and a power ratio of $c_4/c_3$ between both pump beams. The image (at a delay of 2ps) clearly shows the Nile Red while the Lapis Lazuli grain is strongly suppressed. The same can be observed in the time domain data shown in Fig. 3(d). Only the ESA signal is observable (red, ROI2), the GSB process of Lapis (blue, ROI1) is entirely suppressed. To further validate the result of polarization modulation, we also imaged the sample with conventional intensity modulation for parallel and perpendicular polarization. These two measurements can be used to compute the result of polarization modulation (equations (1) & (2)). The computed data are also shown in Fig. 3(d) as green and purple curves, reproducing the measured signal. The result from polarization modulation of the GSB species (Lapis Lazuli) as well as the computed polarization modulation images are shown in the supplementary information. Note that we exclude the time delays in which pump and probe pulse have a temporal overlap, indicated by the yellow shaded area in Fig. 3(b) and Fig. 3(d). During these delays, nonlinear interactions like two photon absorption, cross phase modulation and stimulated Raman scattering are likely to take place and add additional polarization dynamics to the same chemical species. For demonstration purposes, we focus here on the processes with a lifetime only.

While polarization modulation provides increased chemical specificity and imaging speed, there are some challenges. Instead of one pump beam, there are two pump beams, potentially increasing relative laser intensity and beam pointing noise. Also, the total laser power used for imaging is often limited and needs to be divided between the two pump pulse trains, leading to an overall lower signal level and correspondingly lower signal-to-noise ratio (this effect can readily be seen in the increased noise level in Fig. 3(c)).

Instead of modulating the time delay or the relative polarization between two successive pump pulses, we can modulate the pulse length. In our setup this is most easily achieved by inserting dispersive material in one of the two pump beam paths. Temporally short and lengthened pump pulses interact differently with the sample depending on the involved nonlinearity. For example, a two-photon excitation tends to get enhanced more strongly by shorter pump pulses than single-photon excitation. Hence, pulse length modulation pump-probe imaging could be used to enhance (or reduce) chemical contrast that scales with peak pulse intensity, like single-, two-photon-, or three-photon processes. However, convolution with the instrument response function (which is mainly given by the pulse width of pump and probe laser) in the detection makes the actual separation between chemical species less clear than with polarization modulation.

In summary, we have demonstrated new imaging modalities for pump-probe microscopy by a simple experimental upgrade of existing instrumentation. A time-delay modulation scheme suppresses detrimental thermal effects and eliminates unwanted radiative signals by keeping the optical power in the sample constant during image acquisition. We also demonstrated that polarization modulation can selectively image nonlinear processes and therefore provide additional chemical contrast to distinguish different chemical species. We believe that this simple experimental modification provides powerful tools to increase chemical contrast.

**Funding sources.** This project has been made possible in part by grant number 2019-198099 from the Chan Zuckerberg Initiative DAF, an advised fund of Silicon Valley Community Foundation (M.C.F.). This material is also based upon work supported by the National Science Foundation Division of Chemistry under Award No. CHE-1610975 (M.C.F.). D.G. is recipient of a Walter Benjamin fellowship from the German Research Foundation (DFG).

**Acknowledgments**. We thank Jake Lindale, Shannon Eriksson and Kevin Zhou for suggestions, Simone Degan, Chonglin Guan and Yuheng Liao for providing samples.

# Beyond intensity modulation: new approaches to pump-probe microscopy - Supplemental Information

## POLARIZATION DEPENDENCE OF GROUND STATE BLEACHING AND EXCITED STATE ABSORPTION

In pump-probe microscopy, pump and probe laser pulses interact with dipole transitions in the sample. Here, we derive a simple model that describes the effect of laser polarization on the measured pump-probe signal.

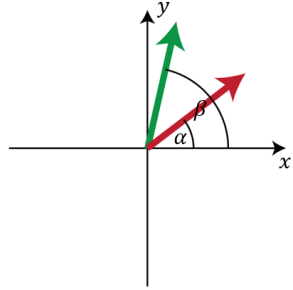

*Figure S1: Polarization state of probe (red arrow) and pump (green arrow) pulse*

We define the state of polarization of the probe pulse to be linear and with angle α and the state of the pump pulse to be linear and with angle β, as indicated in figure S1, with respect to the laboratory reference frame. Let's further assume that the pump pulse interacts with a transition from an electronic ground state to an electronic excited state $|0\rangle \rightarrow |1\rangle$. For excited state absorption events, the probe pulse interacts with a transition from the excited state to a target state $|1\rangle \rightarrow |x\rangle$. In case of a ground state bleaching event, the probe pulse addresses the same transition as the pump pulse $|0\rangle \rightarrow |1\rangle$. The projected orientations of these dipolar transitions in the polarization plane are defined by an azimuthal angle $\gamma_{01}$ and $\gamma_{1x}$, respectively, also with respect to the laboratory reference frame. A measured pump-probe signal of a single molecule is proportional to

$$S \propto (\boldsymbol{\mu_{01}} \cdot \boldsymbol{E_{pu}})^2 (\boldsymbol{\mu_{1x}} \cdot \boldsymbol{E_{pr}})^2 = (\mu_{01}\mu_{1x}E_{pu}E_{pr})^2 \cos^2(\beta - \gamma_{01})\cos^2(\alpha - \gamma_{1x}) \qquad S1$$

with $\boldsymbol{\mu_{01}} = \mu_{01}R(\gamma_{01})\boldsymbol{e_x}$ and $\boldsymbol{\mu_{1x}} = \mu_{1x}R(\gamma_{1x})\boldsymbol{e_x}$ transition dipole moments, $\boldsymbol{E_{pu}} = E_{pu}R(\alpha)\boldsymbol{e_x}$ the electric field of the pump pulse and $\boldsymbol{E_{pr}} = E_{pr}R(\beta)\boldsymbol{e_x}$ the electric field of the probe pulse. $\boldsymbol{e_x}$ is the cartesian unit vector along the x-direction and $R(x)$ is the standard two-dimensional rotation matrix around the angle $x$.

## GROUND STATE BLEACHING

The pump-probe dynamics of Lapis Lazuli are dominated by ground state bleaching (GSB). In a GSB process, pump and probe pulse compete for the same transition, hence $\boldsymbol{\mu_{1x}} = \boldsymbol{\mu_{01}}$. Without loss of generality we choose the probe polarization to be $\alpha = 0$. The above expression simplifies to

$$S_{gsb} \propto (\boldsymbol{\mu_{01}} \cdot \boldsymbol{E_{pu}})^2 (\boldsymbol{\mu_{01}} \cdot \boldsymbol{E_{pr}})^2 = (\mu_{01}\mu_{1x}E_{pu}E_{pr})^2 \cos^2(\beta - \gamma_{01})\cos^2(\gamma_{01}) \qquad S2$$

Within the focal volume of pump and probe laser are typically many molecules and the measured signal is a statistical average over the angular distribution $f(\gamma_{01})$ of the molecules. For simplicity, we model this angular distribution to be $f(\gamma_{01}) = 1$ if $\theta \in [\bar{\gamma} - \delta\gamma, \bar{\gamma} + \delta\gamma]$ within an angular range and 0 elsewhere.

$$\langle S_{gsb} \rangle \propto \int_0^{2\pi} d\theta\, S_{gsb} f(\gamma_{01}) = \int_{\bar{\gamma}-\delta\gamma}^{\bar{\gamma}+\delta\gamma} d\gamma_{01}\, S_{gsb} = a + b\cos[2(\beta - \beta_{gsb})] =: s_{gsb}(\beta) \qquad S3$$

The parameters $a, b$ and $\beta_{gsb}$ are functions of the integration boundaries $\bar{\gamma}$ and $\delta\gamma$. For the case in which the molecules are randomly oriented within the sample ($\delta\gamma = \pi$), we would find $a = 2/3, b = 1/3$ and $\beta_{gsb} = 0$.

## EXCITED STATE ABSORPTION

The pump-probe dynamics of Nile Red are dominated by excited state absorption (ESA). In this case, the pump pulse excites molecules from the ground state to an excited state $|0\rangle \rightarrow |1\rangle$ and the probe pulse further excites the population to a higher state $|1\rangle \rightarrow |2\rangle$. Within the focal volume of pump and probe laser there are typically many molecules and the measured signal is a statistical average over the angular distribution $f(\theta)$ of the molecules. Here, we need an angular distribution $f_1(\theta)$ for the first transition and a second distribution $f_2(\theta)$ for the second transition. We model them in the same way as in the GSB case. The measured ESA signal is the statistical average over both distributions

$$\langle S_{esa} \rangle \propto \int_0^{2\pi} d\gamma_{01} d\gamma_{12} \, S_{gsb} f_1(\gamma_{01}) f_2(\gamma_{12}) = \int_{\overline{\gamma_1}-\delta\gamma_1}^{\overline{\gamma_1}+\delta\gamma_1} d\gamma_{01} \int_{\overline{\gamma_2}-\delta\gamma_2}^{\overline{\gamma_2}+\delta\gamma_2} d\gamma_{12} \, S_{gsb}$$
$$= c + d \cos[2(\beta - \beta_{esa})] := s_{esa}(\beta) \qquad S4$$

The parameters $c, d$ and $\beta_{esa}$ are functions of the integration boundaries $\overline{\gamma_1}, \overline{\gamma_2}, \delta\gamma_1$ and $\delta\gamma_2$. For the case in which the molecules are randomly oriented within the sample ($\delta\gamma_1 = \delta\gamma_2 = \pi$), we would find $c = const$ and $d = 0$.

## DECOMPOSITION OF TRANSIENT ABSORPTION SIGNALS BY POLARIZATION

Our phantom sample is a mixture of Lapis Lazuli and Nile Red and in general, the measured transient absorption signal can be written as superposition of both components:
$$m_{\Delta t}(\beta) \propto g_0(\Delta t) s_{gsb}(\beta) + e_0(\Delta t) s_{esa}(\beta)$$
with $g_0(\Delta t)$ (or $e_0(\Delta t)$) the dependence of GSB (or ESA) signal as a function of time delay $\Delta t$ between pump and probe pulse and with $s_{gsb}(\beta)$ (or $s_{esa}(\beta)$) the polarization dependence of the of GSB (or ESA) as a function of relative polarization angle $\beta$ between probe and pump pulse.

For two measured transient absorption curves with parallel ($\beta = 0$) and perpendicular ($\beta = 90$) polarization between pump and probe we get

$$\begin{pmatrix} m_{\Delta t}(0) \\ m_{\Delta t}(90) \end{pmatrix} = \begin{bmatrix} s_{gsb}(0) & s_{esa}(0) \\ s_{gsb}(90) & s_{esa}(90) \end{bmatrix} \begin{pmatrix} g_0(\Delta t) \\ e_0(\Delta t) \end{pmatrix}. \qquad S5$$

This system of linear equations can be solved for the individual transient absorption signals of GSB $g_0(\Delta t)$ and ESA $e_0(\Delta t)$ by inverting the coefficient matrix

$$g_0(\Delta t) = -\frac{s_{esa}(90) \, m_{\Delta t}(0)}{s_{gsb}(90) s_{esa}(0) - s_{gsb}(0) s_{esa}(90)} + \frac{s_{esa}(0) \, m_{\Delta t}(90)}{s_{gsb}(90) s_{esa}(0) - s_{gsb}(0) s_{esa}(90)} \qquad S6$$

$$e_0(\Delta t) = \frac{s_{gsb}(90) \, m_{\Delta t}(0)}{s_{gsb}(90) s_{esa}(0) - s_{gsb}(0) s_{esa}(90)} - \frac{s_{gsb}(0) \, m_{\Delta t}(90)}{s_{gsb}(90) s_{esa}(0) - s_{gsb}(0) s_{esa}(90)}. \qquad S7$$

The transient absorption curves of the individual components can be reconstructed solely by two measurements with different polarization and by knowledge of the polarization dependence.

# EXPERIMENTAL POLARIZATION CHARACTERIZATION

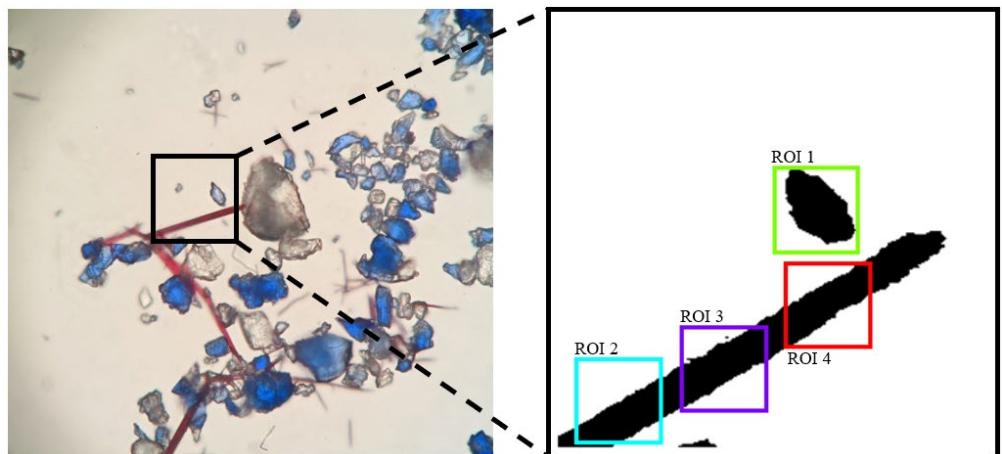

*Figure S2: Brightfield image of sample containing a mix of Nile Red and Lapis Lazuli. The inset represents the field of view used for pump-probe microscopy. The binary image of the inset visualized the 4 regions of interest used for sample calibration.*

## POLARIZATION CHARACTERIZATION OF LAPIS LAZULI

We perform the polarization characterization with a region of interest that predominantly contains Lapis Lazuli (ROI1). The pump-probe microscope is configured for conventional intensity modulation (pump laser is switched on and off) and we image the sample for 7 different polarization angles $\beta \in [0,15,30,45,60,75,90]$ degrees between pump and probe laser. Each measurement is averaged over ROI1. A sum of an instantaneous process and a biexponential decay are fitted to each of the 7 transient absorption curves, convolved with the instrument response function. The instantaneous process is necessary to account for 2-photon absorption, cross-phase modulation and (potentially) stimulated Raman signals, but it is not used for further evaluation.

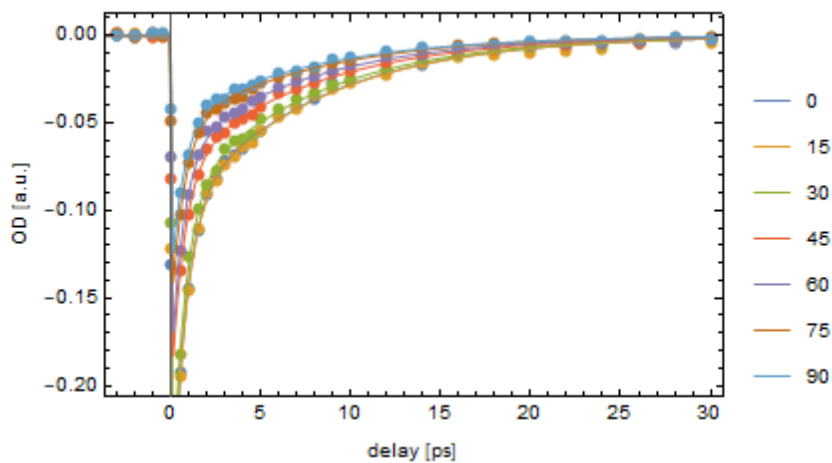

*Figure S3: Transient absorption curves of ROI1 (Lapis Lazuli) measured with conventional intensity modulation for 7 different polarization angles (see color coding) between pump and probe laser pulse. The solid points represent measurement data and the lines represent their corresponding fit.*

The fit result contains two amplitudes (for each exponential decay) per polarization angle which are normalized to their corresponding extremum. Both amplitudes are fitted to equation S3, as shown in figure S4 and result in the polarization dependence of Lapis Lazuli.

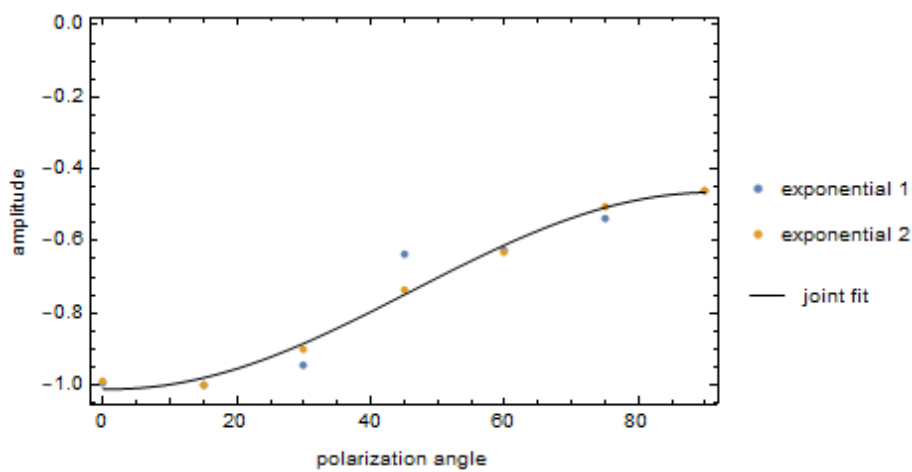

*Figure S4: Fit (solid black line) of exponential decay amplitudes (blue and yellow points) as a function of polarization angle between pump and probe laser pulse*

## POLARIZATION CHARACTERIZATION OF NILE RED

We perform the polarization characterization with 3 regions of interest that predominantly contains Nile Red (ROI2-4). The pump-probe microscope is configured for conventional intensity modulation (pump laser is switched on and off) and we image the sample for 7 different polarization angles $\beta \in [0,15,30,45,60,75,90]$ degrees between pump and probe laser. Each measurement is averaged over its corresponding ROI and a combination of instantaneous process and biexponential decays, convolved with the instrument response function, is fitted to each of the 7 transient absorption curves. The instantaneous process is necessary to account for 2-photon absorption, cross-phase modulation and (potentially) stimulated Raman signals, but it is not used for further evaluation. Here we only show data and fit of ROI2, but ROI3 and ROI4 behave similarly.

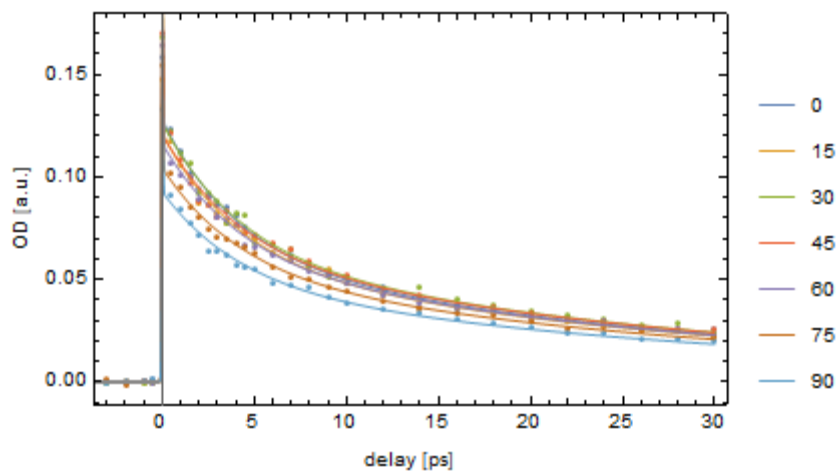

*Figure S5: Transient absorption curves of ROI2 (Nile Red) measured with conventional intensity modulation for 7 different polarization angles (see color coding) between pump and probe laser pulse. The solid points represent measurement data and the lines represent the corresponding fit.*

The fit result contains two amplitudes (for each exponential decay) per polarization angle and per ROI, which are normalized to their corresponding extremum. All 6 amplitudes per angle are fitted to equation S4, as shown in figure S6 and result in the polarization dependence of Nile Red.

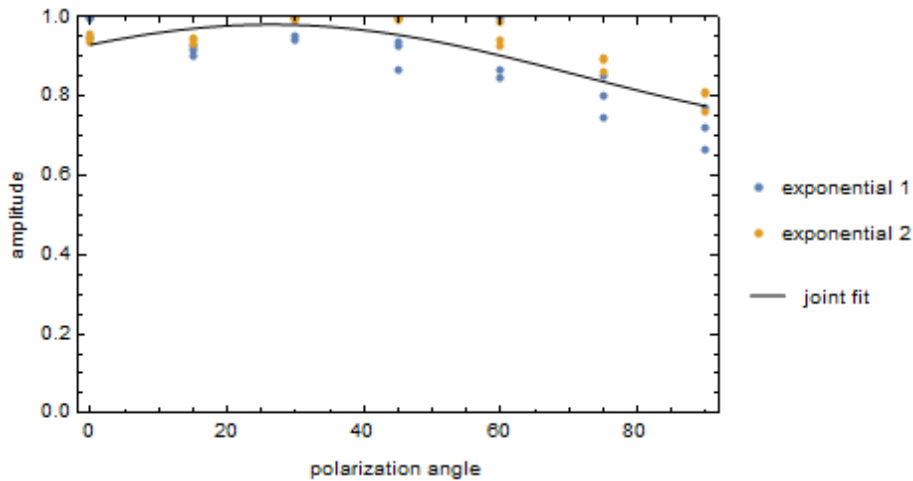

*Figure S6: Fit (solid black line) of exponential decay amplitudes (blue and yellow points) as a function of polarization angle between pump and probe laser pulse*

## PARAMETERS FOR THE DECOMPOSITION OF NONLINEAR PROCESSES

The fit result of the polarization measurements for Lapis Lazuli and Nile Red reveal their polarization dependence to be

$$s_{gsb}(\beta) = -1.01 - 0.46 \cdot Cos[2(\beta - 1.2)]$$

$$s_{esa}(\beta) = 0.85 + 0.13 \cdot Cos[2(\beta - 26.4)].$$

These dependencies allow us to compute or measure the ground state bleaching signal or the excited state absorption signal by usage of equation S6 and S7 or with polarization modulation pump-probe microscopy.

## COMPARISON BETWEEN COMPUTATIONAL DECOMPOSITION AND MEASUREMENT

In the main text we demonstrate how to use polarization-based modulation to selectively suppress nonlinear processes based on their polarization dependence. We can also reach the same result in an indirect way by measuring the same specimen twice with intensity modulation and different polarization angles between pump and probe. The two measurements can be used to compute images solely containing the GSB or ESA species. Here we show the results of polarization-based modulation and compare them to the computed results based on intensity modulation.

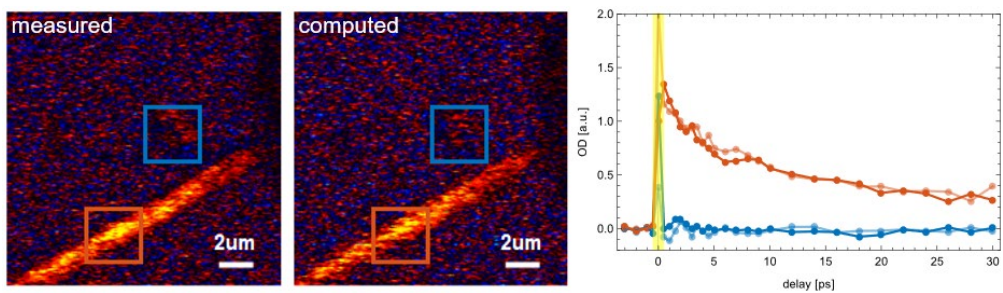

*Figure S7: Measured polarization modulation pump-probe image and intensity modulation based, computed pump-probe image (both at $\Delta t = 2ps$) to suppress the GSB process (blue square). Red (blue) curves represent the transient absorption curve averaged over the ROI indicated by the red (blue) square in the images. Dark colors represent measurements, light colors represent computations.*

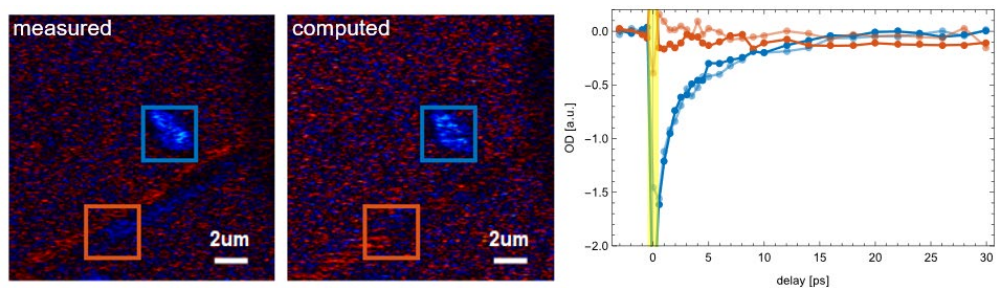

*Figure S8: Measured polarization modulation pump-probe image and intensity modulation based, computed pump-probe image (both at $\Delta t = 2ps$) to suppress the ESA process (red square). Red (blue) curves represent the transient absorption curve averaged over the ROI indicated by the red (blue) square in the images. Dark colors represent measurements, light colors represent computations.*